\documentclass[10pt]{iopart}
\usepackage{epsfig}

\def\Journal#1#2#3#4#5{#1 {\it #2} #3 {\bf #4} #5}
\def\Preprint#1#2{#1 {\it Preprint} #2}
\def\NP{Nucl. Phys.}
\def\PL{Phys. Lett.}
\def\PR{Phys. Rev.}
\def\PRL{Phys. Rev. Lett.}
\def\PRT{Phys. Rep.}
\def\RPP{Rept. Prog. Phys.}

\begin{document}

\title{Effective hadronic Lagrangian for charm mesons}

\author{Ziwei Lin and C.M. Ko}

\address{Cyclotron Institute and Physics Department, Texas A\& M University, 
College Station, TX 77843-3366, USA}

\begin{abstract}
An effective hadronic Lagrangian including the charm mesons is introduced 
to study their interactions in hadronic matter. Using coupling
constants that are determined either empirically or by the SU(4) symmetry,
we have evaluated the absorption cross sections of $J/\psi$ and the scattering 
cross sections of $D$ and $D^*$ by $\pi$ and $\rho$ mesons.
\end{abstract}

\section{Introduction}
Suppression of charmonium production in high energy heavy ion collisions 
is a possible signature for the quark-gluon plasma (QGP) 
formed in ultra-relativistic heavy ion collisions \cite{matsui}. 
Recent data from the Pb+Pb collision at 
$P_{\rm lab}=158$ GeV$/c$ in the NA50 experiment at CERN \cite{na50} have 
shown an anomalously large $J/\psi$ suppression in high $E_{\rm T}$ events.
While there are suggestions that this anomalous suppression may be due to 
the formation of the QGP \cite{qgp1,qgp2}, more 
conventional mechanisms based on $J/\psi$ absorption by comoving 
hadrons have also been proposed as a possible explanation
\cite{conv1,conv2}. Since the latter scenario depends on the value of 
$J/\psi$ absorption cross sections by hadrons, 
it is important to have better knowledge of these cross sections in order
to understand the observed anomalous charmonium suppression 
\cite{reviews}. Even in heavy ion collisions at 
the Relativistic Heavy Ion Collider (RHIC), where the QGP 
is most likely to be formed, the effect of hadronic absorption 
of $J/\psi$ is still non-negligible \cite{zhang}. Furthermore, one needs
to know if $J/\psi$ can also be produced from the hot hadronic matter 
in the later stage of heavy ion collisions \cite{ko,redlich}.

Also, charm mesons such as $D$ and $D^*$ are expected to be abundantly
produced in heavy ion collisions at RHIC energies and beyond.
It has been shown that the high mass ($M > 2$ GeV) dilepton spectrum at RHIC 
may be dominated by decays from charm and bottom hadrons \cite{dilep}. 
Since charm quarks may lose appreciable energies  
in a quark-gluon plasma via gluon radiations,
study of charm meson spectrum in heavy ion 
collisions could provide useful information on the properties of the QGP 
\cite{closs1,closs2}.
However, charm mesons may interact strongly with hadrons during 
later stage of heavy ion collisions, leading to 
possible changes in their final spectra. 
As a result, the dilepton spectrum from the decay of charm meson pairs 
could also be modified \cite{dflow}. To study the energy loss of
charm quarks in the QGP thus requires the understanding of
charm meson interactions in hadronic matter.

Various approaches have been used in evaluating the charmonium absorption
cross sections by hadrons. 
In the quark-exchange model, an earlier study \cite{blaschke} has shown
that the $J/\psi$ absorption cross section by pion, 
$\sigma_{\pi \psi}$, has a peak value of about $7$ mb at
$E_{\rm kin} \equiv \sqrt s-m_\pi-m_\psi \simeq 0.8$ GeV, but a 
recent study \cite{wong} gives a peak value 
of only $\sigma_{\pi \psi} \sim 1$ mb at the same $E_{\rm kin}$ region.  
On the other hand, the perturbative QCD approach \cite{ope} 
predicts that the $J/\psi$ dissociation cross section increases monotonously 
with $E_{\rm kin}$ and has a value of only about $0.1$ mb around 
$E_{\rm kin}\sim 0.8$ GeV. Charmonium absorption cross sections
by hadrons have also been studied in meson-exchange models 
based on effective hadronic Lagrangians. 
Using pseudoscalar-pseudoscalar-vector-meson (PPV) couplings
and without form factors at the interaction vertices,  
Matinyan and M\"uller \cite{muller} have found 
$\sigma_{\pi \psi} \simeq 0.3$ mb at $E_{\rm kin}=0.8$ GeV. 
In a recent study, Haglin \cite{haglin} has included also the 
three-vector-meson (VVV) and four-point couplings (or contact terms),  
and obtained much larger values for the $J/\psi$ absorption cross 
sections. Large discrepancies in the magnitude of $\sigma_{\pi \psi}$ 
(as well as $\sigma_{\rho \psi}$) thus exist among the predictions from
these three approaches. 
In this study, we use a meson-exchange model as in 
\cite{haglin,pv} but treat differently the VVV and four-point couplings 
in the effective Lagrangian. 
For charm meson scattering cross sections with pion and rho meson, 
previous studies from the meson-exchange model include only 
the PPV interactions \cite{ds}.  
We now also use the extended hadronic Lagrangian to study these cross sections.

\section{The effective Lagrangian}

The free Lagrangian for pseudoscalar and vector mesons 
in the limit of SU(4) invariance can be written as
${\cal L}_0= {\rm Tr} \left ( \partial_\mu P^\dagger \partial^\mu P \right )
-{\rm Tr} \left ( F^\dagger_{\mu \nu} F^{\mu \nu} \right )/2$, \linebreak
where $F_{\mu \nu}=\partial_\mu V_\nu-\partial_\nu V_\mu$, 
and $P$ and $V$ denote, respectively, the $4\times 4$ pseudoscalar and 
vector meson matrices in SU(4) \cite{ds}. 
To obtain the couplings between pseudoscalar and vector mesons, 
we introduce the minimal substitution
\begin{equation}
\hspace{-1cm}
\partial_\mu P \rightarrow {\cal D}_\mu P= \partial_\mu P
-\frac{ig}{2} \left [V_\mu P \right ],~
F_{\mu \nu} \rightarrow 
\partial_\mu V_\nu-\partial_\nu V_\mu -\frac{ig}{2} 
\left [ V_\mu, V_\nu \right ].
\end{equation}
The effective Lagrangian is then given by 
\begin{eqnarray}
{\cal L}&=& {\cal L}_0 + ig {\rm Tr} 
\left ( \partial^\mu P \left [P, V_\mu \right ] \right ) 
-\frac{g^2}{4} {\rm Tr} 
\left ( \left [ P, V_\mu \right ]^2 \right ) \nonumber \\
&+& ig {\rm Tr} \left ( \partial^\mu V^\nu \left [V_\mu, V_\nu \right ] 
\right ) 
+\frac{g^2}{8} {\rm Tr} \left ( \left [V_\mu, V_\nu \right ]^2 \right )~.
\label{lagn2}
\end{eqnarray}
Since the SU(4) symmetry is explicitly broken by hadron masses, terms
involving hadron masses are added to Eq.(\ref{lagn2}) using the
experimentally determined values. 

The effective Lagrangian in Eq. (\ref{lagn2}) is
generated by minimal substitution, which is equivalent
to treating vector mesons as gauge particles.  
The gauge invariance leads to the current conservation; i.e., 
in the limit of zero vector meson masses, degenerate pseudoscalar
meson masses, and SU(4) invariant coupling constants, one has
${\cal M}^{\lambda_k \dots \lambda_l}~ p_{j \lambda_j}=0$
for any given process, 
where the index $\lambda_j$ denotes the external vector meson $j$. 
We have checked that all the amplitudes without form factors 
satisfy the requirement of current conservation.  

The above effective Lagrangian allows us to study various 
interactions of charm mesons and $J/\psi$ with hadrons. These
include the charm meson scattering such as 
$\pi D \leftrightarrow \rho D^*$,
the charm meson production and annihilation such as 
$\pi\pi\leftrightarrow D\bar D$, and the charmonium 
absorption and annihilation such as $\pi\psi\leftrightarrow D{\bar D^*}$.
This effective hadronic Lagrangian has also been extended to SU(5) 
to study $\Upsilon$ absorption in hadronic matter \cite{up}.
In the following, we show the results for 
$J/\psi$ absorption and charm meson scattering by pion and rho meson. 

\section{$J/\psi$ absorptions}

For $J/\psi$ absorption by $\pi$ and $\rho$ mesons, 
we study the following processes:
\begin{eqnarray}
\pi \psi \rightarrow D^* \bar D, ~ \pi \psi \rightarrow D \bar {D^*},~
\rho \psi \rightarrow D \bar D, ~ \rho \psi \rightarrow D^* \bar {D^*}. 
\nonumber
\end{eqnarray}
The full amplitudes for these processes can be found in \cite{jpsi}.

From the $D^*$ decay width \cite{dwidth},
the coupling constant $g_{\pi DD^*}$ is found to be $g_{\pi DD^*}=4.4$.
Using the vector meson dominance (VMD) model, we 
determine other three-point coupling constants as \cite{jpsi}
\begin{eqnarray}
g_{\rho DD}=g_{\rho D^* D^*}=2.52~,~
g_{\psi DD}=g_{\psi D^* D^*}=7.64~.  
\end{eqnarray}
For the four-point coupling constants, there is no empirical information,
and we thus use the SU(4) relations to determine their values as 
\begin{eqnarray}
\hspace{-2cm}
g_{\pi \psi DD^*}= g_{\pi DD^*} g_{\psi DD}, ~
g_{\rho \psi D D}= 2~g_{\rho DD} g_{\psi DD}, ~
g_{\rho \psi D^* D^*}=g_{\rho D^* D^*} g_{\psi D^* D^*}.  
\end{eqnarray}

Form factors are introduced at interaction vertices 
to take into account the composite nature of hadrons. 
Unfortunately, there is no empirical information on form factors 
involving charmoniums and charm mesons. 
We thus take the form factors as the usual monopole form 
at the three-point $t$ and $u$ channel vertices, i.e., 
$f_3(t~ {\rm or}~ u)=\Lambda^2 /(\Lambda^2+{\bf q}^2)$,
where $\Lambda$ is a cutoff parameter, and 
${\bf q}^2$ is the squared three momentum transfer in the c.m. frame, 
given by $({\bf p_1}-{\bf p_3})_{\rm c.m.}^2$ 
and $({\bf p_1}-{\bf p_4})_{\rm c.m.}^2$ 
for $t$ and $u$ channel processes, respectively. 
For simplicity, we use the same value for all cutoff parameters, i.e.,
$\Lambda_{\pi D D^*}=\Lambda_{\rho DD}=\Lambda_{\rho D^* D^*}
=\Lambda_{\psi DD}=\Lambda_{\psi D^* D^*}\equiv \Lambda$, 
and choose $\Lambda$ as either $1$ or $2$ GeV 
to study the uncertainties due to form factors.
We also assume that the form factor at four-point vertices is given by 
$f_4=\left [\Lambda^2/(\Lambda^2+\bar {{\bf q}^2}) \right ]^2$,
where $\bar {{\bf q}^2}$ is the average value 
of the squared three momentum transfers in $t$ and $u$ channels,    
$\bar {{\bf q}^2} = p_{i,\rm c.m.}^2+p_{f,\rm c.m.}^2$.

\begin{figure}[tb]
\begin{center}
\begin{tabular}{lr} 
\epsfig{width=0.46 \columnwidth, height=0.36 \columnwidth, file=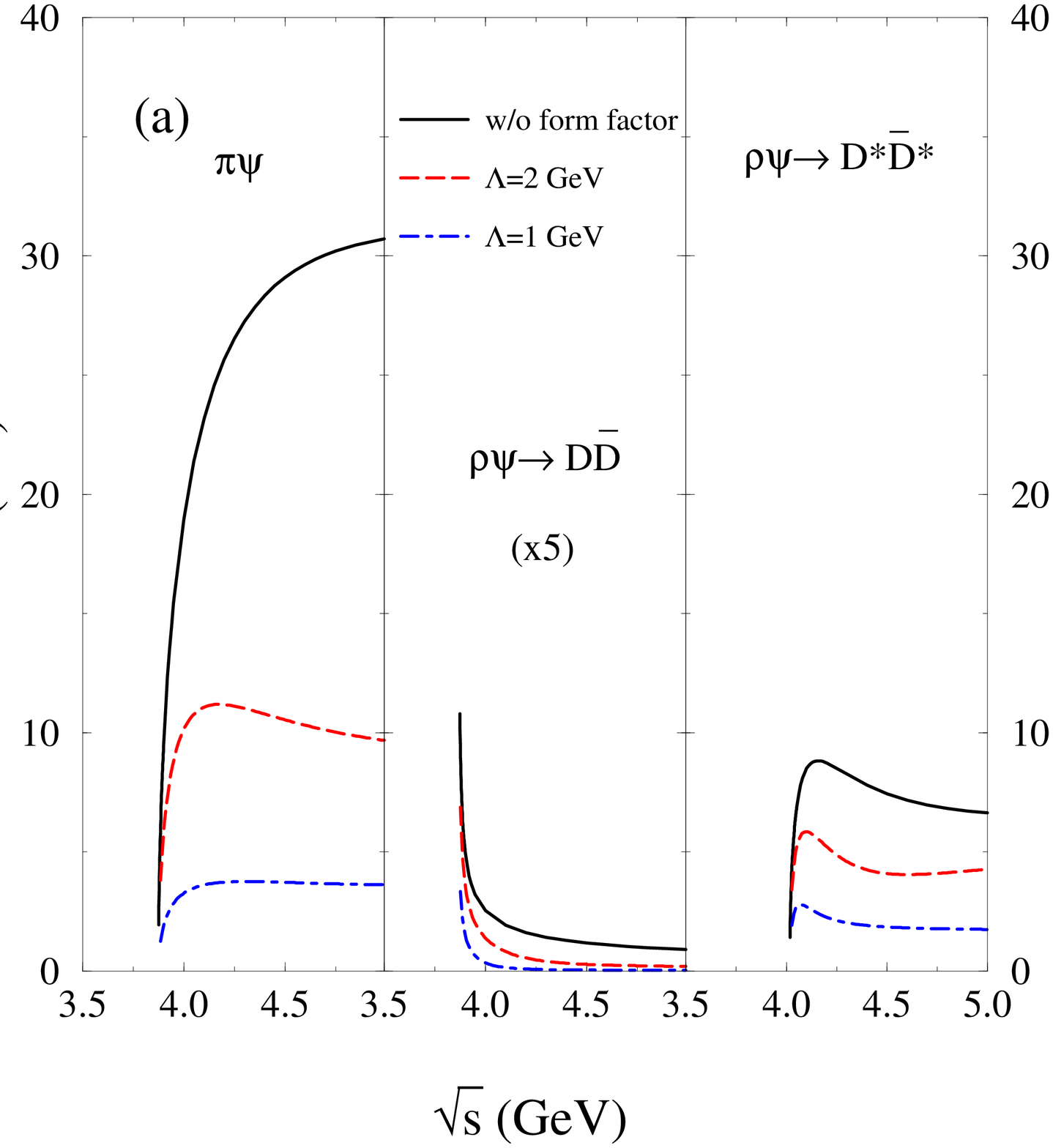}
&
\epsfig{width=0.46 \columnwidth, height=0.36 \columnwidth, file=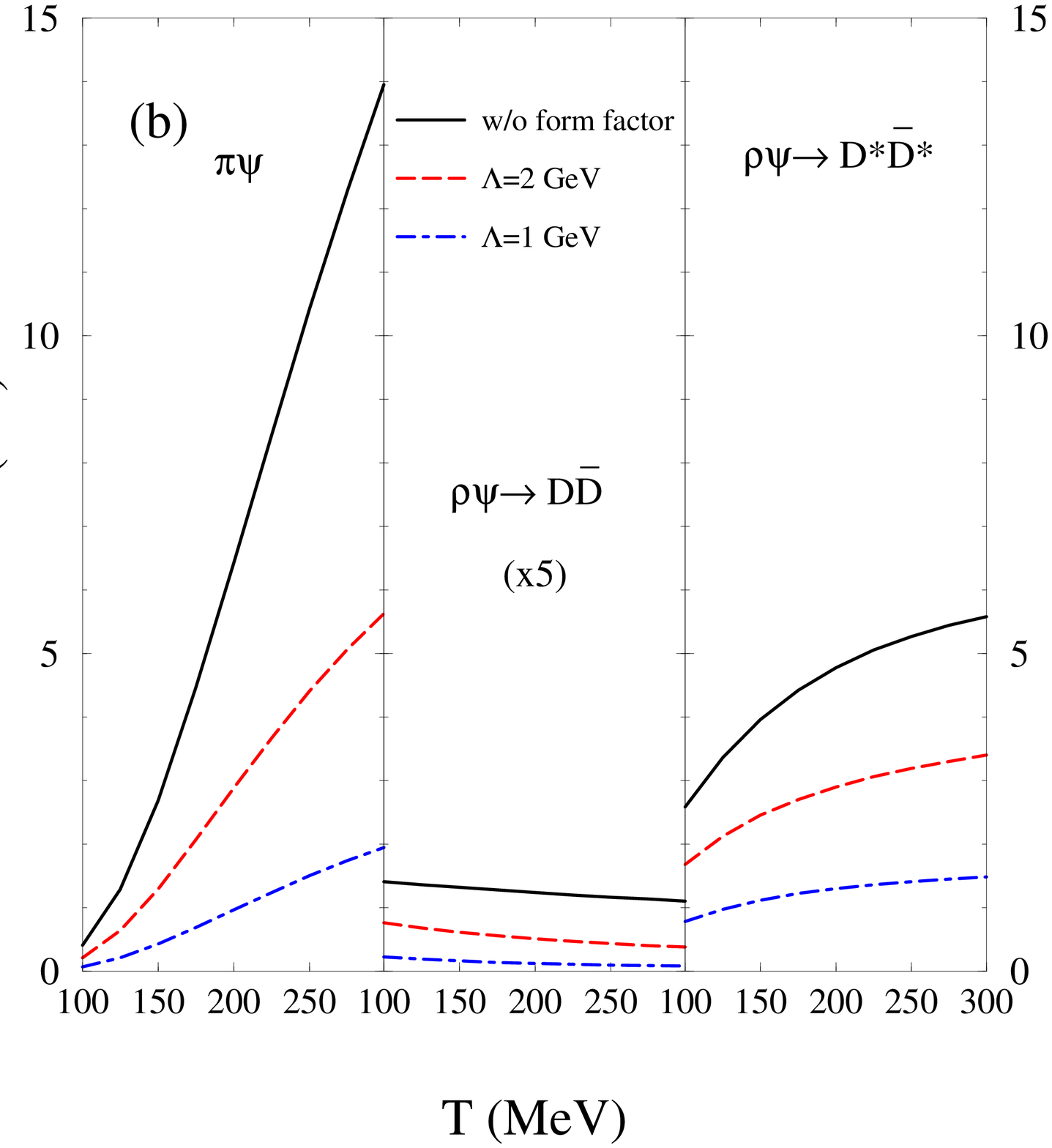}
\end{tabular}
\caption{(a) $J/\psi$ absorption cross sections
as functions of energy.
(b) Thermal-averaged $J/\psi$ absorption cross sections
as functions of temperature.}
\end{center}
\end{figure}

Figure 1a shows the cross section as a function of the c.m. energy 
of initial-state mesons. Although form factors strongly suppress 
the cross sections, the $J/\psi$ absorption 
cross sections remain appreciable after including form factors. 
The values for $\sigma_{\pi \psi}$ and $\sigma_{\rho \psi}$ 
are roughly $7$ and $3$ mb, respectively, 
and are comparable to those used in phenomenological
studies of $J/\psi$ absorption by comoving hadrons in relativistic
heavy ion collisions \cite{conv1,conv2,comover}.
The thermal average of $J/\psi$ absorption cross sections 
is shown in figure 1b. 
At the temperature of $150$ MeV, for example, 
$\langle \sigma_{\pi \psi}v \rangle$ and $\langle \sigma_{\rho \psi}v \rangle$ 
are about $1$ and $2$ mb, respectively. 

\section{Charmed meson scattering}

We consider the following processes 
for charm meson scattering by $\pi$ and $\rho$ mesons:
\begin{eqnarray}
\hspace{-2cm}
\pi D \leftrightarrow \rho D^*,~
\pi D \rightarrow \pi D,~
\pi D^* \rightarrow \pi D^*,~
\pi D^* \leftrightarrow \rho D,~ 
\rho D \rightarrow \rho D,~
\rho D^* \rightarrow \rho D^*. \nonumber
\end{eqnarray}
There are also similar processes for anticharm mesons. 
The full amplitudes for the above processes can be found in \cite{dsg}.

For the coupling constants, 
we use $g_{\rho \pi \pi}=6.1$ \cite{rpp}. 
The SU(4) symmetry then gives $g_{\rho \rho \rho}= g_{\rho \pi \pi}$
and the following relations for the four-point couplings: 
\begin{eqnarray}
\hspace{-2.5cm}
g_{\pi \rho DD^*}= g_{\pi DD^*} g_{\rho DD}, ~
g_{\pi \pi D^* D^*}= 2g_{\pi DD^*}^2, ~
g_{\rho \rho D D}= 2g_{\rho DD}^2, ~
g_{\rho \rho D^* D^*}=g_{\rho D^* D^*}^2.  
\end{eqnarray}

Form factors at the $s$ channel vertices are taken as
$f_3(s)=\Lambda^2/(\Lambda^2+{\bf k}^2)$ \cite{ffs},
with ${\bf k}$ denoting the three momentum of either the incoming or 
outgoing particles in the c.m. frame, i.e., 
${\bf k}^2=p_{i,\rm c.m.}^2$, or $p_{f,\rm c.m.}^2$.    
For form factors at four-point vertices, we take them to be
$f_4^\prime=\left [ \Lambda^2/(\Lambda^2+\bar {{\bf k}^2}) \right ]^2$, 
where $\bar {{\bf k}^2}$ is the average value 
of the squared three momenta in the form factors 
for the $s$, $t$, and $u$ channels, i.e., 
$\bar {{\bf k}^2}=5({p_{i,\rm c.m.}^2+p_{f,\rm c.m.}^2})/6$.

\begin{figure}
\begin{center}
\begin{tabular}{lr} 
\hspace{-0.55cm}
\epsfig{width=0.48 \columnwidth, height=0.46 \columnwidth, file=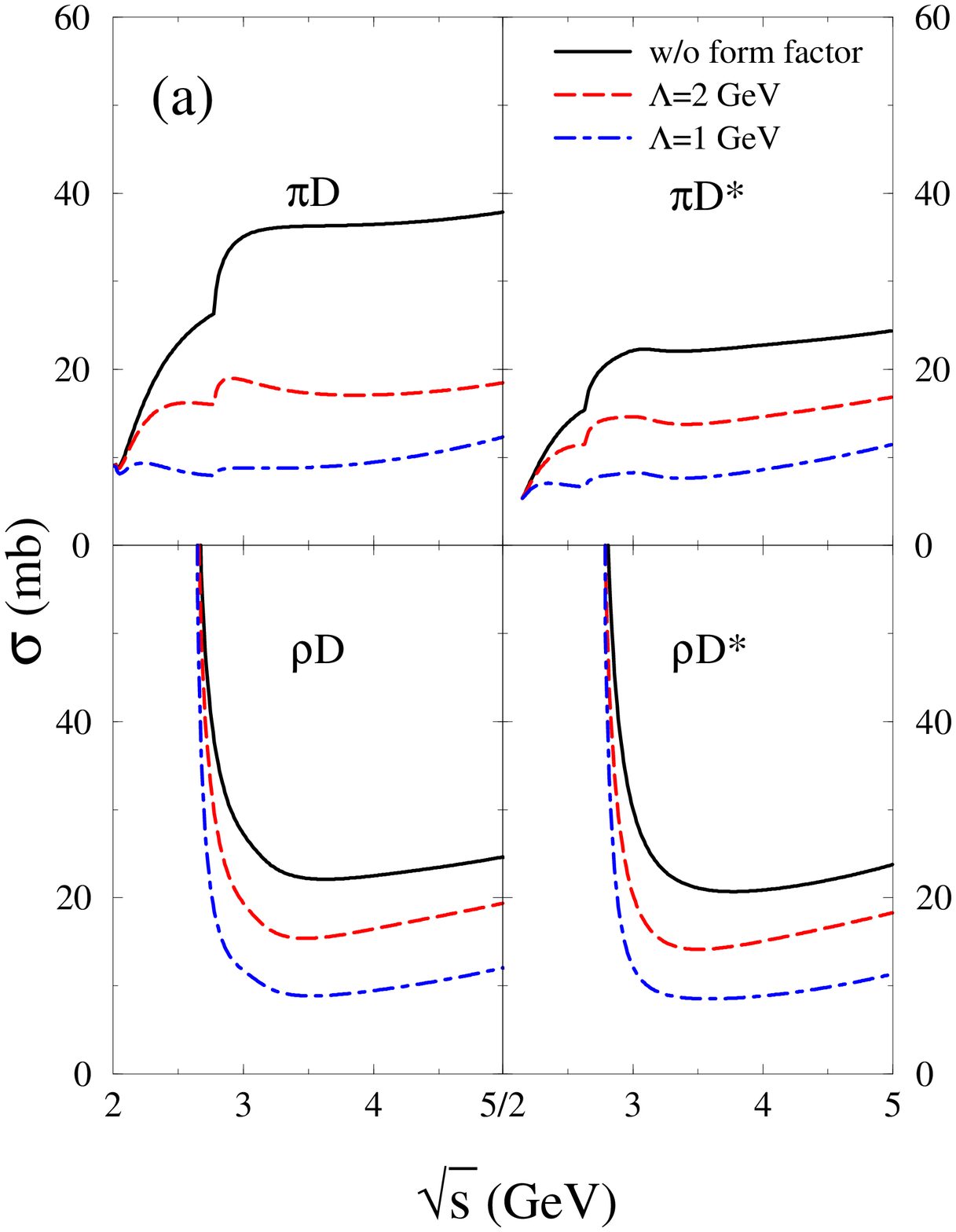}
&
\epsfig{width=0.48 \columnwidth, height=0.46 \columnwidth, file=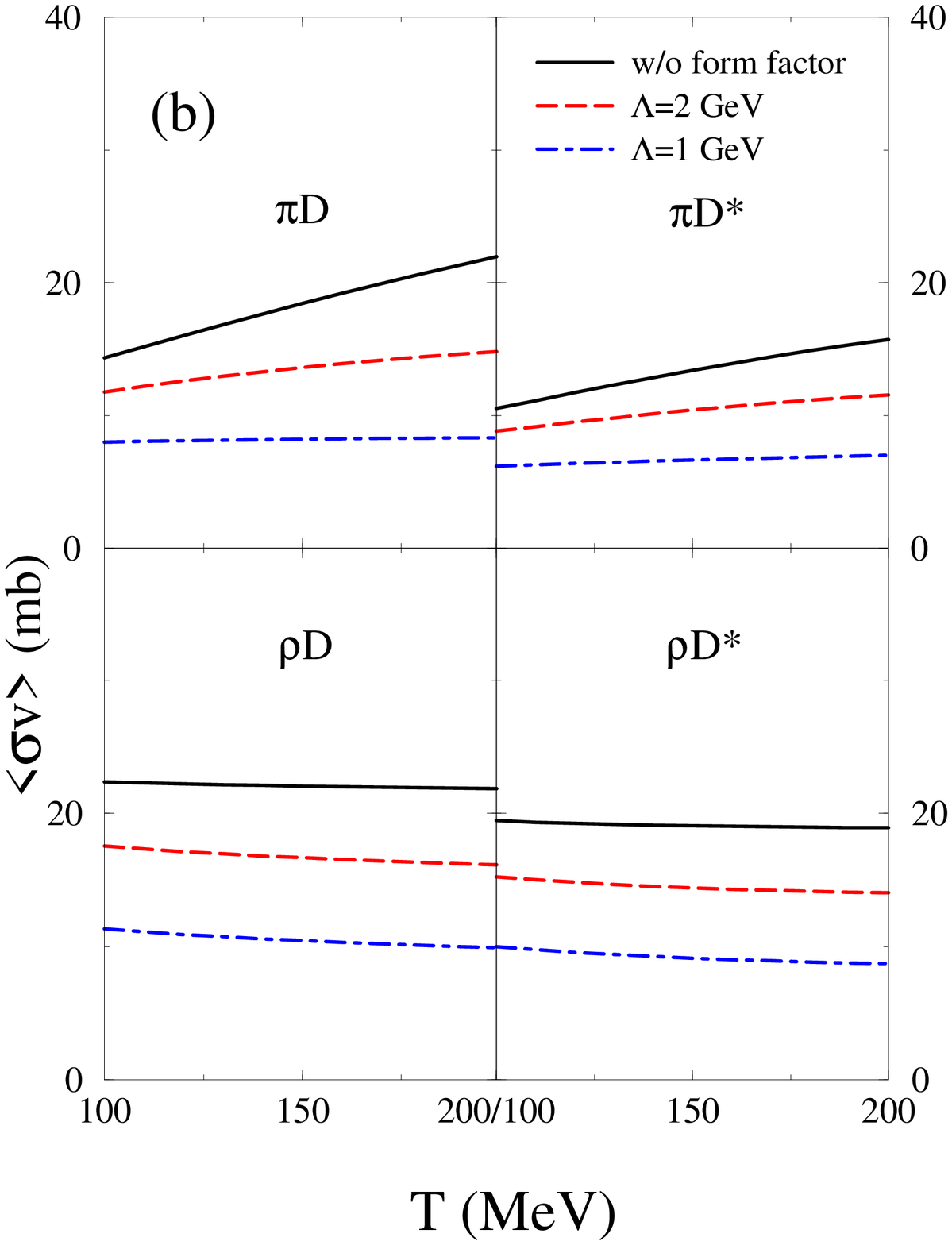}
\end{tabular}
\caption{(a) Scattering cross sections of $D$ and $D^*$  
as functions of energy. 
(b) Thermal-averaged cross sections
as functions of temperature.}
\end{center}
\end{figure}

The charm meson scattering cross sections and their thermal averages 
are shown in figure 2a and 2b, respectively. 
As expected, the magnitude of the cross sections decreases with decreasing 
cutoff parameter. For the cutoff parameters used here, the 
cross sections for $\pi D, \pi D^*, \rho D$ and $\rho D^*$ scattering 
are all roughly between 10 and 20 mb.
We note that form factors only suppress modestly 
(by a factor of two and less) these total cross sections 
and their thermal averages.
This is due to the fact that these cross sections are dominated by elastic 
processes, which involve small momentum transfer near the threshold. 
In contrast, the process $\pi \psi \rightarrow D^* \bar D$ 
has a large threshold, and form factors suppress its cross section
by as much as a factor of 8. 

\section{Summary}

In summary, we have introduced an effective hadronic Lagrangian, that 
includes the charm mesons, to study the interactions of 
charmed mesons and $J/\psi$ in hadronic matter. In particular, 
we have calculated the absorption cross sections of $J/\psi$ 
and scattering cross sections of charmed mesons by $\pi$ and $\rho$ mesons.  
We find that the $J/\psi$ absorption cross sections are
much larger than those in a previous study, 
where only pseudoscalar-pseudoscalar-vector-meson couplings were considered.
Including form factors at the interaction vertices, 
the values for $\sigma_{\pi \psi}$ and $\sigma_{\rho \psi}$ 
are about $7$ mb and $3$ mb, respectively,  
and their thermal averages at the temperature of $150$ MeV 
are roughly $1$ mb and $2$ mb, respectively.   
These values suggest that the absorption of $J/\psi$ by comoving hadrons 
may play an important role in $J/\psi$ suppression 
in relativistic heavy ion collisions. 
We also find that the scattering cross sections of $D$ or $D^*$ by 
$\pi$ or $\rho$ mesons are all about $10$ mb and 
thus expect these scatterings to significantly modify the charm meson 
spectra in heavy ion collisions. 

\section*{Acknowledgments} 
This work was supported in part by the National Science Foundation under 
Grant No. PHY-9870038, the Welch Foundation under Grant No. A-1358,
and the Texas Advanced Research Program under Grant No. FY99-010366-0081.

{}

\end{document}